\documentclass[prl,twocolumn,showpacs,amsmath,amssymb,nofootinbib]{revtex4}
\usepackage[dvips]{graphicx}
\usepackage{enumerate}
\newcommand \beq{\begin{eqnarray}}
\newcommand \eeq{\end{eqnarray}}

\begin{document}

\title{Spectral continuity in dense QCD}

\author{Tetsuo Hatsuda$^{1}$, Motoi Tachibana$^{2}$, Naoki Yamamoto$^{1}$} 
\affiliation{
$^{1}$Department of Physics, The University of Tokyo, Tokyo 113-0033, Japan\\
$^{2}$Department of Physics, Saga University, Saga 840-8502, Japan}

\begin{abstract}

The vector mesons
 in three-flavor quark matter  with chiral and diquark condensates 
 are studied  using the in-medium QCD sum rules.
  The diquark condensate leads  to a mass splitting between
   the flavor-octet and flavor-singlet channels. 
   At high density,
   the singlet vector meson disappears from the low-energy spectrum, while
 the octet vector mesons survive as  
  light excitations with a mass comparable to the fermion gap.
  A possible connection between the light gluonic modes and the flavor-octet
   vector mesons at high density is also discussed.
\end{abstract}

\pacs{12.38.Lg,21.65.Qr}

\maketitle

There are three characteristic phases in  quantum chromodynamics (QCD) at finite 
temperature ($T$) and baryon chemical potential ($\mu_{\rm B}$): 
the hadronic phase at low  $T$ and $\mu_{\rm B}$,  
the quark-gluon plasma phase at high $T$ and $\mu_{\rm B}$  \cite{YHM}, and the 
 color superconductivity (CSC) phase  at low $T$ and high $\mu_{\rm B}$ \cite{CSC}.
   In recent years,  numbers of theoretical and experimental progresses have been
    made for understanding the  
  thermal (classical) phase transition at high $T$  between the hadronic phase and
  the quark-gluon plasma phase. On the other hand,
   the nonthermal (quantum) phase transition
    at low $T$  between the hadronic phase and the CSC phase
 such as the meson condensed phase, the crystalline  
 Fulde-Ferrell-Larkin-Ovchinikov phase, gluon condensed phase, etc. 
  \cite{CSC} has not been
 fully understood in spite of its relevance to the interior of neutron stars \cite{NS}.

   In determining the phase structure at finite density,
   interplay between the dynamical breaking of chiral symmetry 
   due to quark$-$anti-quark pairing, $\langle \bar{q}q \rangle $,
     and  color superconductivity 
   due to quark-quark pairing, $\langle qq \rangle $, may play a crucial role
    \cite{kitazawa02,highT}.\footnote{
   Another important interplay would be induced by 
   confinement characterized by the Polyakov loop \cite{PNJL} 
    which we will not discuss in this paper.}
       Recently,  it was suggested that such an 
    interplay
     between $\langle \bar q q \rangle$ and $\langle qq \rangle$
     induced by the QCD axial anomaly 
    leads to a  smooth crossover between the hadronic phase
   and the CSC phase \cite{HTYB06}.
Furthermore, it was shown in Ref.~\cite{YTHB07} that the ordinary pion
associated with dynamical chiral symmetry breaking at low density
and the generalized pion associated with the color superconductivity at
high density
are mixed with each other in the crossover region: They form
a light pionic mode satisfying a generalized Gell-Mann$-$Oakes$-$Renner
relation in the first order of the quark mass $m_q$;
\beq
\label{eq:gGOR}
f_{\pi}^2 m_{\pi}^2 = m_q \left(- \langle \bar{q}q \rangle +
\Gamma \langle qq \rangle^2 \right) .
\eeq
Here $f_{\pi}$, $m_{\pi}$ and $\Gamma$
are the generalized pion decay constant, the pion mass, and a parameter
characterizing the strength of the $U(1)_A$ axial anomaly, respectively.
In the low density limit with vanishing diquark condensate,
Eq.~(\ref{eq:gGOR}) reduces to the usual Gell-Mann$-$Oakes$-$Renner relation,
$f_{\pi}^2 m_{\pi}^2=-m_q \langle \bar qq \rangle$ \cite{GOR}.
On the other hand, at intermediate to high densities, where
the diquark condensate dominates over the chiral condensate,
the pion mass may be dictated by the strength of the axial anomaly,
$f_{\pi}^2 m_{\pi}^2 \sim m_q \Gamma \langle qq \rangle^2$.
At asymptotic high densities where $\Gamma$ is suppressed,
the $O(m_q^2)$-term not considered in the above
formula eventually takes over to lead
$f_{\pi}^2m_{\pi}^2 \propto m_q^2 \langle qq \rangle^2$ \cite{ARW, SS00}.
This is an explicit realization of the ``spectral continuity of
hadrons" from low density to high density and may have close
relation to the idea of hadron-quark continuity in dense QCD \cite{SW99}.

  The purpose of this paper is to  present a novel attempt
  for studying the spectral continuity across the crossover region
  on the basis of the QCD sum rules
   \cite{SVZ79} and its in-medium generalization \cite{HKL93}. 
  In our approach,   the resonance  parameters in dense matter are related to the 
   chiral and diquark condensates through gauge invariant
    correlation functions.
  In the following, we will focus on the flavor octet and singlet
   vector mesons  (such as $\rho,\omega, \phi, K^*$)
  in three-flavor quark matter at zero temperature 
  with massless  u, d, s quarks.  A smooth crossover 
  in the intermediate density region  is assumed according to \cite{HTYB06,YTHB07}.

Let us start with the vector-current correlation function in quark matter:
\beq
\label{eq:r-corr}
\Pi_{\mu \nu}^{AB} (\omega, {\vec q})  =  {i}
\int d^4 x\ {\rm e}^{iqx} \langle {\rm R} J_{\mu} ^{A} (x)J_{\nu} ^{B} (0) \rangle ,
\eeq
 where R denotes the retarded product, $q^{\mu}=(\omega, \vec{q})$,
 $J_\mu ^{A}  = \bar q\tau ^{A} \gamma _\mu  q$ with  $q=$(u,d,s).  
 The flavor $U(3)$ generators  
 $\tau^{A}\ (A=0,...,8)$ are normalized as tr$[\tau^{A} \tau^{B}]=2\delta^{AB}$. 
  The expectation value of an operator ${\cal O}$  
  with respect to the quark matter is denoted as 
   $\langle {\cal O} \rangle$.
 One can decompose Eq.(\ref{eq:r-corr}) 
  by introducing the longitudinal and transverse 
  invariants ($\Pi^{AB}_{\rm L,T}$) as
$\Pi^{AB}_{00}={\vec q}^{\ 2} \Pi^{AB}_{\rm L}$ and 
$\Pi^{AB}_{ij}=(\delta_{ij}-q_i q_j/{\vec q}^{\ 2})\Pi^{AB}_{\rm T} 
+ q_i q_j \omega^2 \Pi^{AB}_{\rm L}/{\vec q}^{\ 2}$.
 Since there is no distinction between the transverse and longitudinal
  components in the limit ${\vec q}=0$, 
   it is sufficient to consider 
   $\Pi^{AB}_{\rm L}(\omega) \equiv \Pi^{AB}_{\rm L}(\omega, {\vec 0})$
 in treating the vector mesons at rest in quark matter \cite{HKL93}.
  Also, it is convenient to define
 the flavor-octet and singlet correlation functions 
  because of the flavor $SU(3)$ symmetry: 
 $\Pi_{\rm L}^{(8)} \equiv \frac{1}{8} \sum_{A = 1}^{8} {\Pi_{\rm L}^{AA} }$
 and $\Pi_{\rm L}^{(1)}\equiv \Pi_{\rm L}^{00}$. 
 
 The dispersion relation for  $\Pi_{\rm L}$ reads
\beq
\label{eq:dispersion}
 \Pi_{\rm L}(\omega) 
 = \int_0^\infty  \frac{\rho(u)}{u^2 - (\omega + i \epsilon)^2}du^2 
 - ({\rm subtraction}),
\eeq
  where the spectral function is 
  given by $\rho(u)=(1/\pi){\rm Im}  \Pi_{\rm L}(u)$. 
  In QCD sum rules, $\Pi_{\rm L}(\omega)$
  in the deep Euclidean region ($Q^2 \equiv -\omega^2 \rightarrow \infty$) 
  is evaluated with the operator product expansion (OPE)
  and is compared with the dispersion integral 
  in Eq.~(\ref{eq:dispersion}) under suitable parametrization of $\rho(u)$.
  Not only the Lorentz scalar operators but also
  the tensor operators in OPE contibute to the correlation functions
  in the medium \cite{HKL93}.
  In the massless quark matter with finite quark chemical potential $\mu$, 
  the matrix element of an operator with canonical dimension $d$
  has the general form  
\beq
\langle {\cal O} \rangle  = f(\alpha_{\rm s}) \mu^d
  +\langle \! \langle {\cal O} \rangle \! \rangle,
\eeq
 where
  the first term is  calculable order by order 
  in $\alpha_{\rm s}$, while the second term 
  contains genuine nonperturbative
  effects  such as the chiral and diquark condensates.  
     
  To show how our approach works, we start with 
  noninteracting  quark matter with the quark 
  chemical potential $\mu$. The OPE for $\Pi_{\rm L}^{(8,1)}$
  up to $O(1/Q^6)$ in this case reads \cite{HKL93}
\beq
\label{eq:free-ope}
\Pi_{\rm L}^{\rm (free)}(Q)
= - \frac{1}{2\pi ^2 } \ln Q^2 +
 \frac{16}{9}     \frac{\langle {q^{\dagger}i \partial_{0}   q} \rangle}{Q^4}
+ \frac{64}{9} \frac{\langle {q^{\dagger}i \partial_{0}^3 q} \rangle}{Q^6} . 
\eeq
 The matrix elements of the twist 2 quark operators in Eq.~(\ref{eq:free-ope})
 can be evaluated as 
\beq
\left\langle {q}^{\dagger}i \partial_{0 }^n q \right\rangle 
 = (-i)^{n-1} \frac{ 9\mu^{n+3} } {(n+3)\pi^2}.
\eeq
The asymptotic expansion Eq.~(\ref{eq:free-ope}) 
 matches exactly the spectral function in the free quark matter as shown in Fig.~1(a):
\beq
\label{eq:free-spectral-function}
\rho^{\rm (free)}(u)=F \delta(u^2)+A\theta(u^2-S_0),
\eeq
  with the resonance parameters, $S_0=(2\mu)^2$, $A = 1/(2\pi^2)$, and $F=A S_0$. 
  The pole part  at zero energy corresponds to the scattering  of 
  quarks on the Fermi surface with the external current, i.e., the
  Landau-damping term.  The continuum part corresponds to the 
  decay of the external current into quark-antiquark pair
  with the Pauli blocking effect. 
  The corrections of the form 
  $\alpha_{s} \ln Q^2$ and $\alpha_{\rm s} (\mu/Q)^n$
   to Eq.~(\ref{eq:free-ope})
  can be taken into account in perturbation theory 
  and are compensated by the perturbative corrections to the 
   resonance parameters and the spectral  shape in  
   Eq.~(\ref{eq:free-spectral-function}).

 \begin{figure}[h]
\begin{center}
\includegraphics[width=6.0cm]{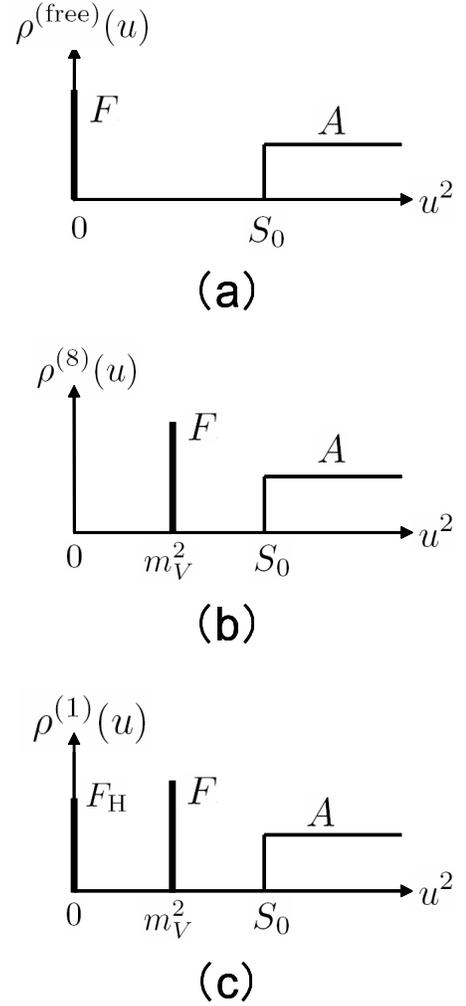}
\end{center}
\vspace{-0.5cm}
\caption{The spectral function for the correlation of vector currents in dense quark matter:
 (a) The case without non-perturbative condensates, 
 (b) the case for octet correlation  with  chiral and diquark condensates,
 and (c) the case for singlet correlation with chiral and diquark condensates.
}
\label{fig:spect}
\end{figure}

\begin{figure}[h]
\begin{center}
\includegraphics[width=4.0cm]{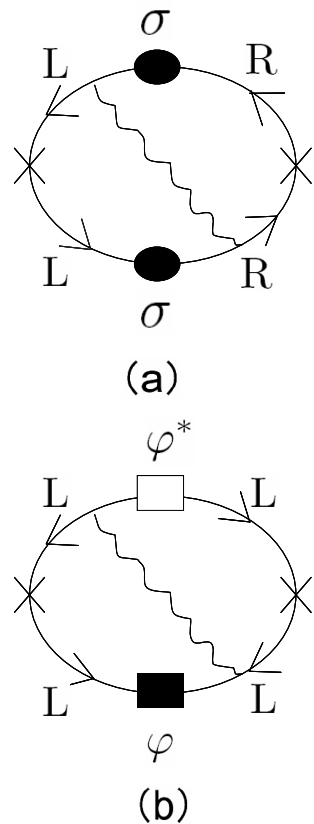}
\end{center}
\vspace{-0.5cm}
\caption{Examples of $\delta \Pi_{\rm L}^{(8,1)}$ from (a) the chiral condensate and (b) 
 the diquark condensate.
Quarks are denoted by solid lines with their chirality structure (L and R), while
 the gluon is denoted by a wavy line.
the chiral condensate $\sigma$ by a black circle
and the diquark condensate $\varphi$ ($\varphi^*$) 
by a black (white) square. }
\label{fig:4quark}
\end{figure}

 Let us now study the effect of 
 ``nonperturbative" condensates.
 They introduce extra four-quark terms
  in Eq.~(\ref{eq:free-ope}) \cite{HKL93,other-op}
 \beq
\label{eq:ope}
\! \! \! \! \! \! \! \! \! \! \!   \Pi_{\rm L}^{(8,1)}\! \! \! \! &=& \Pi_{\rm L}^{\rm (free)}
 + \delta \Pi_{\rm L}^{(8,1)}, \\
\label{eq:octet-4}
\! \! \! \! \! \! \! \! \! \!  \delta \Pi_{\rm L}^{(8)} \! \! \! &=& \! \! \! - \frac{\pi \alpha_s}{Q^6} \! \!
\left[\frac{1}{4} \langle \! \langle( {\bar q\gamma _\mu  \gamma _5 \tau ^a \lambda ^{a'} q} )^2 \rangle \! \rangle  
\! +\! \frac{8}{27} \langle \! \langle ( {\bar q\gamma _\mu  \lambda ^{a'} q} )^2 \rangle \! \rangle   
\!  \right] \! \! ,  \\
\label{eq:singlet-4}
\! \! \! \! \! \! \! \! \! \!  \delta \Pi_{\rm L}^{(1)} \! \! \! &=& \! \! \! - \frac{\pi \alpha_s}{Q^6} \! \!
\left[   { 2\langle \! \langle( {\bar q\gamma _\mu  \gamma _5 \tau ^0 \lambda ^{a'} q} )^2 \rangle \! \rangle  } 
\! +\! \frac{8}{27} {\langle \! \langle( {\bar q\gamma _\mu  \lambda ^{a'} q} )^2 \rangle \! \rangle  }\! \right] 
\! \! , 
\eeq
 where $\tau^{a=1 \sim 8} $ ($ \lambda^{a'=1 \sim 8} $)
  are the flavor (color) $SU(3)$ generators, and
  the repeated indices  are summed over from 1 to 8.
  The four-quark condensates in  Eqs.~(\ref{eq:octet-4}) and (\ref{eq:singlet-4})
  induce two independent corrections to the 
  current correlation function 
  as illustrated in Fig.~\ref{fig:4quark} for 
  $\langle \! \langle 
  ({\bar q\gamma _\mu  \gamma _5 \tau ^A \lambda ^{a'} q} )^2 \rangle \! \rangle $.
  The graph in  Fig.~\ref{fig:4quark}(a) corresponds to
    $\bar{q}q$ pairing which connects the vector currents with opposite chiralities,
   while that  in  Fig.~\ref{fig:4quark}(b) corresponds to  $qq$ pairing
   which connects the vector currents with the same chirality.
  Here we consider the most attractive channels in each case, namely
   $\bar{q}q$ pairing in the color-flavor singlet and spin-parity $0^+$ channel,
  and  $qq$ pairing in the color-flavor locked (CFL) and  spin-parity $0^+$ channel \cite{ARW},
\beq
 \langle \! \langle  \bar q_{i}^{\alpha} q_{j}^{\alpha} \rangle \! \rangle &=& {\rm diag}(\sigma,\sigma,\sigma), \\
 \frac{1}{4} \epsilon_{ijk} \epsilon_{\alpha\beta\gamma} 
\langle \! \langle q_{j}^{\beta} C \gamma_5 \Lambda_+ q_{k}^{\gamma} \rangle \! \rangle 
&=& {\rm diag} (\varphi,\varphi,\varphi).
 \eeq
  Here  $i, j, k$ ($\alpha, \beta, \gamma)$ are the flavor (color) indices and 
   $\Lambda_+$ is the projection operator
 to the positive energy quarks.
 After rewriting the four-quark operators in
  Eqs.~(\ref{eq:octet-4}) and (\ref{eq:singlet-4}) in the chiral basis
  and making the appropriate Fierz rearrangement together with the factorization ansatz  \cite{fact}
\beq
 \langle \! \langle {\cal O} \rangle \! \rangle =
  \sum_l \langle \! \langle {\cal P}_l \cdot {\cal P}_l  \rangle \! \rangle 
  \simeq
    \sum_l \langle \! \langle {\cal P}_l  \rangle \! \rangle^2,
\eeq
we obtain
\beq
\! \! \! \! \delta \Pi_{\rm L}^{(8,1)}\! \! \! 
 & \simeq & \Pi_{\sigma} + \Pi_{\varphi}^{(8,1)},  \\
\label{eq:four-quark}
\label{eq:barqq-part}
\Pi_{\sigma}  &=& 
- \frac{448 \pi \alpha_s}{81Q^6} \sigma^2 , \\
\label{eq:octet-qq-part}
\Pi_{\varphi}^{(8)}  \! \! &=&
-\frac{5}{22}\Pi_{\varphi}^{(1)}
= - \frac{320\pi \alpha_s}{27Q^6} \varphi^2  .
\eeq 
  Since the chiral condensate is flavor-diagonal, it does not 
  distinguish  between octet and singlet.
  On the other hand, the diquark condensate has color-flavor structure 
  which can differentiate the flavor structure in
  Eqs.~(\ref{eq:octet-4}) and (\ref{eq:singlet-4}).
  This is the reason why the 
  flavor-octet and flavor-singlet vector mesons, which are almost
  degenerate at low density, tend to split at high density 
  as will be shown below.  

Let us  first examine  the flavor-octet vector mesons
 under the influence of the nonperturbative condensates.
 Since there is a gap at the Fermi surface in the color superconducting
 quark matter, we do not expect the Landau-damping contribution,
 in contrast to Eq.~(\ref{eq:free-spectral-function}). Instead
  we may have flavor-octet collective modes with finite mass. 
  With these observations, we introduce the following ansatz for the 
  spectral function as shown in Fig.~1(b):
\beq
\label{eq:octet-spectral-function}
\rho^{(8)}(u)=F\delta(u^2-m_{\rm V}^{2})+A\theta(u^2-S_0).
\eeq
Here $m_{\rm V}$ and $S_0$ are
the vector meson mass and the continuum threshold, respectively. 
 
 Substituting the ansatz Eq.~(\ref{eq:octet-spectral-function})
  into Eq.~(\ref{eq:dispersion}), carrying out the asymptotic expansion 
  in terms of $1/Q^2$ after subtraction,
   and comparing the result with the OPE expression in Eq.~(\ref{eq:ope}),
   we end up with the following finite energy sum rules: 
\beq
\label{eq:octet-fesr2}
F &-& A S_0 = 0,  \\ 
\label{eq:octet-fesr3}
2Fm_{\rm V}^2  &-& A S_0^2 =  - \frac{{(2\mu )^4 }}{{2\pi ^2 }},
\\ 
\label{eq:octet-fesr4}
3Fm_{\rm V}^4  &-& A S_0^3 =  - \frac{{(2\mu )^6 }}{{2\pi ^2 }} 
+ \langle \! \langle {\cal O}^{(8)} \rangle \! \rangle , 
\eeq
with  $A=1/(2\pi^2)$ and 
\beq
\langle \! \langle {\cal O}^{(8)} \rangle \! \rangle  
=-\frac{448}{27}\pi \alpha_s \left(\sigma^2  + \frac{15}{7}  \varphi^2 \right).
\eeq
   When $\mu=0$ and $\varphi=0$, the solution of these equations
  reduces to the standard formula \cite{KPT83}:
\beq
\label{eq:vacuum-octet-mass}
\left( m_{\rm V}^{(8)} \right)^2 
\xrightarrow[\mu \rightarrow 0]{ }
\left(\frac{448\pi^3 \alpha_s}{27} \sigma^2 \right)^{\frac{1}{3}}.
\eeq
 When $\mu \neq 0 $ with both
  $\sigma$ and $\varphi$ finite,  we obtain the quartic equation for $S_0$
 from Eqs.~(\ref{eq:octet-fesr2}-\ref{eq:octet-fesr4}),
\beq
\label{eq:quartic}
t^4+6t^2-4(1+r^{(8)})t-3=0,
\eeq
where $t$ and $r^{(8)}$ are dimensionless parameters defined as 
$t=S_0/(2\mu)^2$ and
 $r^{(8)}\equiv-2\pi^2 \langle \! \langle {\cal O}^{(8)} \rangle \! \rangle /(2\mu)^6$, respectively.
In a situation  where $0 \le r^{(8)} \ll 1$,  we have a unique solution,  
\beq
\label{eq:octet-mass}
 \left( m_{\rm V}^{(8)} \right)^2  &\simeq& 
\frac{56{\pi ^3 \alpha _s }}{81{\mu ^4 }}
\left( { \sigma^2   
+ \frac{15}{{7}} \varphi^2 } \right), \\
S_0 &\simeq& (2\mu)^2 + (m_{\rm V}^{(8)})^2, \\
 F &=& \frac{S_0}{2\pi^2}.
\eeq
This is a new formula 
relating the mass of octet vector mesons to the chiral and diquark condensates.
Since $0\le r^{(8)} < 0.1$ is satisfied as long as 
 $(|\sigma|^{1/3},|\varphi|^{1/3})/\mu < 0.3/\alpha_{\rm s}^{1/6}$,
  the small $r^{(8)}$ approximation is safe
 for all densities expected in quark matter.
 The characteristic value of the flavor-octet vector meson mass
 is 
   $m_V^{(8)} \simeq 100 {\rm MeV}$ for  $\mu=500$ MeV, $\alpha_s \sim 1$ and 
 $\sigma \sim \varphi \sim (150 {\rm MeV})^3$.

At asymptotic high density, 
we have $\sigma \sim 0$ and 
the weak coupling relation \cite{S00},
\beq
\varphi= \frac{{3} {\mu^2} \Delta}{\pi \sqrt{2\pi \alpha_s}},
\eeq 
with the fermion gap $\Delta$. Then we obtain
\beq  
m_{\rm V}^{(8)} 
 \rightarrow   \sqrt{\frac{20}{3}} \ \Delta,
\eeq
which  implies the 
existence of light vector mesons in the flavor-octet channel  
 in the high density limit \cite{2delta}. 
 A similar result was also obtained in \cite{RSWZ00,JS04}.
 
 The above discussion leads to a question about the fate of the flavor-singlet 
 vector meson. 
 In this channel, unlike the vector octet, the Nambu-Goldstone scalar (referred to as H) 
 associated with the spontaneous breaking of $U(1)_{\rm B}$ symmetry
 contributes to the spectral function \cite{FI05}.
 (This is analogous to the situation where not only the $a_1$ meson but also
   the pion contribute   to the  axial-vector current correlation in the vacuum.)
 Then we should modify the ansatz for the spectral function 
 Eq.~(\ref{eq:octet-spectral-function}) 
as
\beq
\label{eq:singlet-spectral-function}
\rho^{(1)}(u)=F_{\rm H} \delta(u^2)+ F\delta(u^2-m_{\rm V}^{2})+A\theta(u^2-S_0),
\eeq
This situation is shown in Fig.~1(c).
The finite energy sum rules  in the singlet channel are obtained 
 by the
 replacements: $F \rightarrow F+F_{\rm H}$ 
 in Eq.~(\ref{eq:octet-fesr2}) and 
\beq 
\langle \! \langle {\cal O}^{(8)} \rangle \! \rangle \rightarrow
 \langle \! \langle {\cal O}^{(1)} \rangle \! \rangle 
  = -\frac{448}{27}\pi \alpha_s \left( \sigma^2- \frac{66}{7} \varphi^2 \right)
\eeq
  in Eq.~(\ref{eq:octet-fesr4}).
Although  the sum rules are not closed due to
 the extra parameter $F_{\rm H}$, we can
 still  select a possible solution under physical constraints
  such as the positivity of the spectral function and
  the small magnitude  of   $r^{(1)}
  \equiv -2\pi^2 \langle \! \langle {\cal O}^{(1)} \rangle \! \rangle /(2\mu)^6$. 
 It is remarkable here that unlike the case of $r^{(8)}$, the parameter $r^{(1)}$
  could change its sign depending on the relative magnitude of the condensates
  $\sigma$ and $\varphi$.
  
  For  $ r^{(1)} > 0$ with fixed $F$,
    we find two solutions: a heavy one with 
    $S_0 >  \left( m_{\rm V}^{(1)} \right)^2 > 2\mu^2$ and a
     light one,
\beq
\label{eq:singlet-mass}
 \left( m_{\rm V}^{(1)} \right)^2  \simeq  
\frac{1}{f} \frac{56{\pi ^3 \alpha _s }}{81{\mu ^4 }}
\left( {\sigma^2   
- \frac{66}{{7}} \varphi^2 } \right),   
\eeq
with $f \equiv 2 \pi^2F/(2\mu)^2$. 
 Since  the mass in
 Eq.~(\ref{eq:singlet-mass})
  is essentially determined by the QCD condensates, 
  we adopt this light solution 
  as the physical one  connected continuously to the 
  flavor-singlet vector meson at low density. 
  
  For $r^{(1)} <  0$ with fixed $F$, 
 we have only a  heavy solution with $S_0 >  \left( m_{\rm V}^{(1)} \right)^2 > (2\mu)^2$.
  Since the mass in this case is above the threshold for $\bar{q}q$ decay,
   we do not expect it  to appear as a sharp resonance in reality.
 This indicates that
 the light flavor-singlet vector meson can exist only
  below a critical chemical potential $\mu_{\rm c}$ 
  determined by the condition, $\sigma^2 =   {\frac {66}{7}} \varphi^2$.
 The value of $\mu_{\rm c}$ cannot be estimated without
   knowing the precise $\mu$ dependence of $\sigma$ and $\varphi$ which is
   beyond the ability of the QCD sum rules.
    However, as long as 
 $|\sigma|$ ($|\varphi|$) is a decreasing (increasing)
  function of the chemical potential, $\mu_{\rm c}$ is located 
  below the crossover point at $|\sigma| = |\varphi|$. 
  
  Note here that, for sufficiently high density with $|r^{(1)}| \ll 1$ with fixed $F$,
 one can show  a relation 
\beq
F_{\rm H} \rightarrow  \frac{2\mu^2}{\pi^2}
\eeq
 from our sum rules.
 This is consistent with the known $F_{\rm H}$ estimated in
  the weak coupling calculation at high density \cite{SS00, FI05}.

\begin{figure}[t]
\begin{center}
\includegraphics[width=8.5cm]{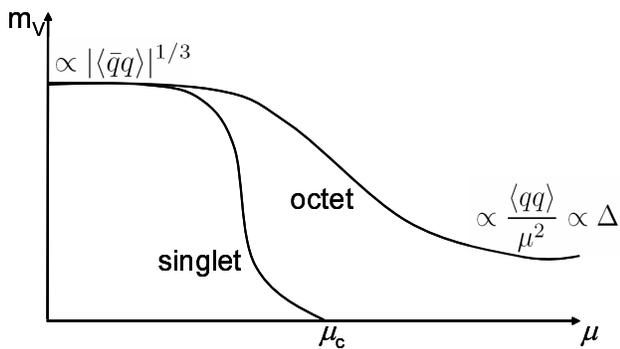}
\end{center}
\vspace{-0.5cm}
\caption{Schematic plot of the masses of flavor-octet and flavor-singlet 
 vector mesons as a function of $\mu$.
  The magnitude of the chiral condensate $\sigma$ (the diquark condesate $\varphi$) 
   is assumed to decrease (increase) monotonically as $\mu$ increases. }
\label{fig:vector_mass}
\end{figure}

 On the basis of our analysis, 
 we draw a schematic plot of the masses of
 the octet and singlet vector mesons 
 as a function of the chemical potential $\mu$  in Fig.~\ref{fig:vector_mass}.
 At low density, vector mesons   are almost 
  degenerate and form a nonet structure.  
 As the density increases, the effect of the diquark condensate becomes
  prominent: It  contributes positively (negatively) to 
  $m_{\rm V}^{(8)}$ ($m_{\rm V}^{(1)}$)
     as seen from Eqs.~(\ref{eq:octet-mass}) and (\ref{eq:singlet-mass}), and
     increases their mass splitting.
  At higher density, 
  the octet vector mesons survive as 
  light excitations,
   while the singlet vector meson disappears from the low-energy spectrum.

 Remembering the fact that  flavors and colors are mixed
  in the CFL phase,  flavor-octet vector mesons at high density
   cannot be distinguished  from the color-octet gluons.
  This suggests that 
  the light gluonic modes
   with the mass of  $O(\Delta)$  (the light CFL plasmon)  
 found in  \cite{CGN01}
   has a close connection to our octet vector mesons  
     at asymptotic high density.
       
In summary, we have studied the vector mesons
  in quark matter  with massless u, d, s quarks by using
 the  in-medium QCD sum rules.
  We have derived formulas relating   
  vector meson masses to the chiral and diquark condensates.
  Owing to  the different role played by the chiral and diquark condensates,
   we find spectral continuity (discontinuity)
  in the flavor-octet (flavor-singlet) channel.  In particular,  
    the octet vector mesons survive at high density as  
  light excitations with a mass comparable to the fermion gap.
  
    Our sum-rule approach provides a novel tool to analyze 
  not only the vector mesons but also the other hadrons in dense matter.
   For example,   the spectral continuity of
    flavored baryons at low density and colored quarks at high density
    is one of the  interesting problems to be examined.
  Also, it is important to study how 
  the strange quark mass whose main effect in OPE comes from the operator
  $m_s \bar{s}s$,  affects  the spectral continuity found in massless quark matter.

 We thank G. Baym, K. Fukushima and C. Sasaki for useful
 comments. N. Y. is supported by the Japan Society for the Promotion of Science
for Young Scientists. T. H. is supported in part by the Grants-in-Aid of the
Japanese Ministry of Education, Culture, Sports, Science, and Technology
(No.~18540253).



\begin{thebibliography}{99}

\bibitem{YHM} Reviewed in, 
 K. Yagi, T. Hatsuda and Y. Miake, {\em Quark-Gluon Plasma},
 Cambridge Univ. press (Cambridge, 2005).

\bibitem{CSC}  Reviewed in, 
M. G. Alford, A. Schmitt, K. Rajagopal and T. Sch\"{a}fer, 
 Rev. Mod. Phys. (in press):  arXiv:0709.4635 [hep-ph]

\bibitem{NS} Reviewed in, 
H. Heiselberg and  V. Pandharipande,
 Ann. Rev. Nucl. Part. Sci. {\bf 50}, 481 (2000).

\bibitem{kitazawa02}
 See e.g.,  M.~Kitazawa, T.~Koide, T.~Kunihiro and Y.~Nemoto,
  Prog.\ Theor.\ Phys.\  {\bf 108}, 929 (2002);
 M.~Buballa, Phys.\ Rept.\  {\bf 407}, 205 (2005).

\bibitem{highT}
 Similar interplay between 
  antiferromagnetism and high $T_c$ superconductivity is seen
 in copper oxide superconductors: See, e.g.,
Q. Chen, J. Stajic, S. Tan and K. Levin, Phys. Rep. {\bf 412}, 1 (2005).

\bibitem{PNJL}
 K. Fukushima, Phys. Lett. B {\bf 591}, 277 (2004);
 S.~Roessner, T.~Hell, C.~Ratti and W.~Weise,  arXiv:0712.3152 [hep-ph].

\bibitem{HTYB06}
T. Hatsuda, M. Tachibana, N. Yamamoto and G. Baym, 
Phys. Rev. Lett. {\bf 97}, 122001 (2006).

\bibitem{YTHB07}
N. Yamamoto, M. Tachibana, T. Hatsuda and G. Baym, 
Phys. Rev. D {\bf 76}, 074001 (2007).

\bibitem{GOR}
M.~Gell-Mann, R.~J.~Oakes and B.~Renner,
Phys.\ Rev.\  {\bf 175}, 2195 (1968).

\bibitem{ARW} 
M. G. Alford, K. Rajagopal and F. Wilczek, Nucl. Phys. B {\bf 537}, 443 (1999).

\bibitem{SS00}
D.~T.~Son and M.~A.~Stephanov, Phys.\ Rev.\ D {\bf 61}, 074012 (2000),
 [erratum] ibid. D {\bf 62}, 059902 (2000). 

\bibitem{SW99} 
T. Sch\"{a}fer and F. Wilczek, Phys. Rev. Lett. {\bf 82}, 3956 (1999);
 K. Fukushima, Phys. Rev. D {\bf 70}, 094014 (2004).


\bibitem{SVZ79}
A. Shifman, A. I. Vainshtein and V. I. Zakharov, Nucl. Phys. B {\bf 147},
385 (1979), ibid. 448 (1979).

\bibitem{HKL93}
T. Hatsuda and S. H. Lee,  Phys. Rev. {\bf C46}, R34 (1992):
T. Hatsuda, Y. Koike and S. H. Lee, Nucl. Phys. B {\bf 394},  221 (1993);
T. Hatsuda, S. H. Lee and H. Shiomi, Phys. Rev. {\bf C52}, 3364 (1995). 
 

\bibitem{other-op}
  There are several operators neglected in Eq.~(\ref{eq:ope}) up to $O(1/Q^6)$,
  whose explicit forms are given in \cite{HKL93}.
  Among others, the nonperturbative gluon condensate
  $\langle \! \langle {\frac{{\alpha _s }}{\pi }G^2 } \rangle \! \rangle$,
   if it survives in 
  quark matter, affects the octet and singlet vector
  mesons in the same way. The nonscalar operators such as the 
  quark-gluon mixed operators and the twist-4 quark operators do
  not produce the chiral or diquark condensates and lead
 only  to perturbative corrections 
 to $\Pi_{\rm L}^{\rm (free)}$.


\bibitem{fact}
 The qualitative conclusions in the present paper do not depend on the factorization
  ansatz.  As for the violation of the 
   factorization in the vacuum, see S. Narison, Phys. Lett. {\bf B624}, 223 (2005).



\bibitem{KPT83}
N. V. Krasnikov, A. A. Pivovarov and N. N. Tavkhelidze, 
Z. Phys. C {\bf 19},  301 (1983).

\bibitem{S00}
T. Sch\"{a}fer, Nucl. Phys. B {\bf 575} 269 (2000).

\bibitem{2delta}
 The precise value of the dimensionless 
 ratio $m_{\rm V }^{(8)}/\Delta$ would be sensitive 
 to the ansatz of the spectral function in our
 approach.

\bibitem{RSWZ00}
M. Rho, E. V. Shuryak, A. Wirzba and I. Zahed,
Nucl. Phys. A {\bf 676}, 273 (2000). 

\bibitem{JS04}
A. D. Jackson and F. Sannino,
Phys. Lett. B {\bf 578}, 133 (2004). 


\bibitem{FI05}
K. Fukushima and K. Iida,
Phys. Rev. D {\bf 71}, 074011 (2005).


 \bibitem{CGN01}
R. Casalbuoni, R. Gatto and G. Nardulli,
Phys. Lett. B {\bf 498}, 179 (2001);
V. P. Gusynin and I. A. Shovkovy,
Nucl. Phys. A {\bf 700}, 577 (2002);
H. Malekzadeh and D. H. Rischke,
Phys. Rev. D {\bf73}, 114006 (2006).

\end{thebibliography}
\end{document}